\documentclass[prb,twocolumn,showpacs,superscriptaddress]{revtex4}

\usepackage{amsmath,amssymb,graphicx,color,bm,epstopdf}

\begin{document}

\title{Magnetic Field Control of the Optical Spin Hall Effect}

\author{S. Morina}
\affiliation{Division of Physics and Applied Physics, Nanyang Technological University 637371,  Singapore}
\affiliation{Science Institute, University of Iceland, Dunhagi 3, IS-107, Reykjavik, Iceland}

\author{T. C. H. Liew}
\affiliation{Division of Physics and Applied Physics, Nanyang Technological University 637371, Singapore}
\author{I. A. Shelykh}
\affiliation{Division of Physics and Applied Physics, Nanyang Technological University 637371, Singapore}
\affiliation{Science Institute, University of Iceland, Dunhagi 3, IS-107, Reykjavik, Iceland}

\begin{abstract}
We investigate theoretically the effect of an external magnetic field on polarization patterns appearing in quantum microcavities due to the optical spin Hall effect (OSHE). We show that increase of the magnetic field perpendicular to the plane of the cavity resulting in the increase of the Zeeman splitting leads to the transition from azimuthal separation of polarizations to their radial separation. This effect can be straightforwardly detected experimentally.
\end{abstract}


\maketitle


\section{Introduction}

The spin Hall effect \cite{DyakonovPerel1971} (SHE) provides a fundamental mechanism for the generation and separation of electron spins, with potential for emerging solid-state spintronic technologies. In the original prediction by Dyakonov and Perel, a spin current is induced perpendicular to the electric current arising from elastic spin- dependent scattering of the carriers on impurities present in a semiconductor. The effect was studied in recent years after its rediscovery by Hirsch \cite{Hirsch1999}, and now there are two distinguishable versions, namely the extrinsic SHE, the one mentioned above, and the intrinsic SHE, which stems from the  the action of the effective magnetic fields produced by Spin Orbit Interaction (SOI) of Dresselhaus or Rashba types on spin of moving electrons or holes \cite{Murakami2003,Sinova2004}.

A challenge for spintronic devices based on electron spins is the rapid dephasing and decay of electron spin currents caused by strong electron scattering \cite{Mishchenko2004,Sih2005}. This motivated the development of an optical analog of spintronics, known now as \textit{spinoptronics} \cite{Shelykh2004}. In this domain, spin currents are produced by 
exciton polaritons which are the elementary excitations of semiconductor microcavities within the strong coupling regime (see e.g., Ref. \onlinecite{Sanvitto2012, Deveaud2007, Kavokin2007}). Being a mixture of quantum well excitons and cavity photons, they possess numerous peculiar properties which distinguish them from other quasiparticles in mesoscopic systems. From the point of view of their spin structure, polaritons are analogical to electrons and can be considered as a two-level system, where  the two allowed spin projections $\pm1$ correspond to the two opposite circular polarizations of the cavity photons coupled with bright excitons.

As for any two-level system, one can introduce the pseudospin vector \textbf{S} for the description of the polarization dynamics of polaritons (see Ref. \onlinecite{ShelykhKavokin2010} for the details). In full analogy with the case of electrons, \textbf{S} undergoes a precession caused by effective magnetic fields, arising from intrinsic or extrinsic polarization splittings. As polaritons contain a photonic component, their decoherence times are orders of magnitude bigger than the decoherence times for the electrons and polariton spin currents can propagate ballistically at distances of tens and hundreds of microns \cite{Langbein2007,Leyder2007,Kammann2012,Wertz2010,Wertz2012}.

Among the magnetic fields acting on the polariton pseudospin is the one provided by longitudinal- transverse (LT) splitting \cite{ShelykhKavokin2010}. It appears due to the combination of the LT splitting for moving an exciton provided by long- range exchange interaction between an electron and a hole forming it \cite{Maialle1993} and transverse electric-transverse magnetic splitting of the photonic cavity mode \cite{Panzarini1999}. In analogy to effective magnetic fields of Rashba and Dresselhaus types for electrons the magnitude of this field is dependent on the wavenumber of the polariton \textbf{k} and vanishes at $k\rightarrow0$. This field provides a major mechanism of polariton spin relaxation in the regime of incoherent pumping \cite{KKavokin2004}, and under coherent excitation can lead to the remarkable phenomenon of the optical spin Hall effect (OSHE), which represents an optical analog of the intrinsic spin Hall effect for electrons \cite{Kavokin2005}.

In the original configuration of OSHE the formation of polarization domains in the far- field emission was provided by the elastic scattering\cite{Langbein2002} of polaritons injected into system by disorder and subsequent rotation of their pseudospins by effective TE- TM magnetic field \cite{Kavokin2005} whose value is given by

\begin{eqnarray}
\label{HLT}
{\mathbf{H}_{LT}}\left( \mathbf{k} \right) &=& {\Delta _{LT}}\left( \mathbf{k} \right){\left( {\cos 2\phi ,\sin 2\phi } \right)^T}\\
\label{DeltaLT}
{\Delta _{LT}}\left( \mathbf{k} \right) &=& {E_{T}}\left( \mathbf{k} \right) - {E_{L}}\left( \mathbf{k} \right)
\end{eqnarray}
Here we measure $\textbf{H}$ in energy units and $E_{T}$ and $E_{L}$
are the dispersion relations of the longitudinally and transverse polarized polaritons, respectively.
Differently from the effective magnetic field provided by Rashba and Dresselhaus SOI, the LT field makes a double angle with the X- axis in the reciprocal space. For a polariton propagating without scattering, with a given
value of $\textbf{k}$, the pseudospin $\mathbf{S}$ precesses around the vector of the effective field, which causes beatings in the polarization components of the emitted light. 

The redistribution of the polaritons in reciprocal space
provoked e.g., by impurity scattering leads to the change of the direction of the rotation of pseudospin and can result in formation of polariton spin currents in the plane of the microcavity.  If the flux of $X$-polarized particles with a pseudospin $\mathbf{S}_0=S_x \mathbf{e}_x$ and wavevector $\textbf{k}=k\textbf{e}_x$ hits an impurity that redistributes the particles over the elastic circle, the rotation of pseudospin of the scattered particles strongly depends on rotation angle. Indeed, the precession amplitude is maximal when $\mathbf{S}\bot\mathbf{H}_{LT}$, i.e. for scattering angles $45^\circ, 135^\circ, 225^\circ, 315^\circ$,  while no precession at all occurs for scattering angles of $0^\circ, 90^\circ, 180^\circ, 270^\circ$ when $\mathbf{S}\|\mathbf{H}_{LT}$. This leads to the appearence
of alternating circularly polarized domains in the four quarters of the XY plane in reciprocal space \cite{Kavokin2005,Leyder2007,Kammann2012}, which can be detected by a far- field emission measurement.

Instead of using disorder scattering, it is also possible to use
a pump spot focused in real space to excite the localized polariton wavepacket \cite{Langbein2007,Kammann2012, Amo2009}. The resulting polaritons are then propagating radially outward from the spot rotating their pseudospins, which leads to the formation of alternating domains in real space, which can be detected in a near- field emission measurement. This configuration is particularly promising for the possibility of investigating nonlinear regimes 
where disorder can be screened \cite{Kammann2012,Flayac2013a,Flayac2013b}.

While it is understood that the optical component of exciton-polaritons is chiefly responsible for providing the main source of longitudinal-transverse splitting and allowing the observation of OSHE in a pure photonic cavity\cite{Maragkou2011}, the excitonic component allows sensitivity to an applied external magnetic field. Magnetic fields have been predicted to effectuate the control of polaritonic effects, such as Josephson oscillations\cite{Zhang2011} and quantum blockade \cite{Zhang2009}. As well, it was demonstrated theoretically that the interplay of TE- TM and Zeeman splittings can lead to the manifestation of Berry phase effects in 1D polariton interferometers \cite{Shelykh2009}

Here, we study the effect of an applied magnetic field on the polariton spin currents and polarization patterns generated by the OSHE. The applied magnetic field, oriented perpendicular to the cavity plane (in the growth direction) induces a Zeeman splitting between the polariton states with circular polarizations and thus provides an additional mechanism of pseudospin rotation. We consider the geometries of single and multiple scattering of polaritons by a disorder potential (original OSHE configuration) and excitation with an optical pump beam of finite spot size where effects of disorder can be neglected. We show that in both cases the increase of the magnetic field leads to the crossover from azimuthal separation of polarizations to radial separation of polarizations.

\section{OSHE provided by disorder scattering}

We start our consideration from the case of the original configuration of OSHE, where pumping was performed by a spatially homogenous coherent linearly polarized beam with well defined momentum \textbf{k} and formation of polarization domains was due to the disorder scattering. To describe a realistic situation we should account for the possible effects of multiple scattering. However, to clearly understand the situation qualitatively, we start with a  hypothetical case of single scattering, for which a semiclassical analytical theory can be developed. Such kind of theory was first built in the original theoretical proposal of OSHE \cite{Kavokin2005} and here we develop it further to take Zeeman splitting into account.

\begin{figure}
\centering
\includegraphics[width=3.41082in]{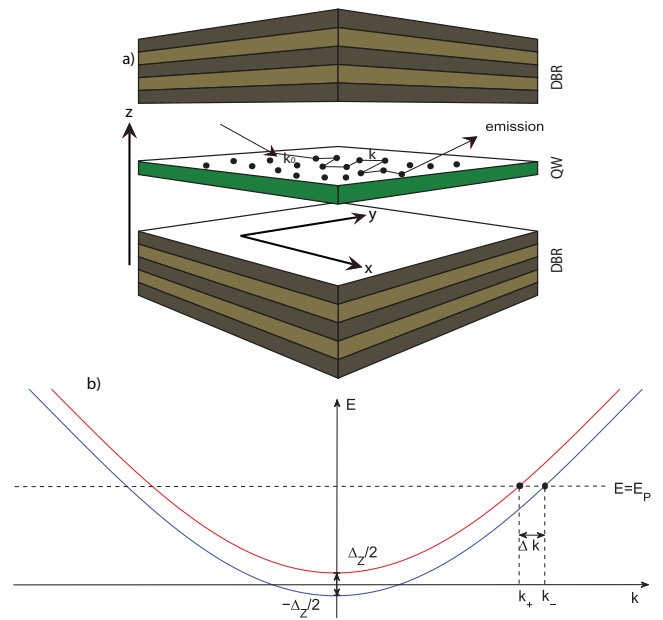}
\caption{(a) Polaritons are excited in the state $\textbf k_0$ by a linearly polarized light directed in the x direction. Their pseudospin vector rotates around a total effective magnetic field, whose direction and magnitude is determined by the angle of the scattered state $\textbf k$ and the external magnetic field which lies along the z direction. (b) A real magnetic field induces Zeeman splitting $\Delta_Z$ between left and right circularly polarized polaritons. Polaritons excited by a CW pump with a definite energy $E_P$ have different wave-numbers for different states of circular polarization.  }
\label{scheme}
\end{figure}

\subsection{Single scattering}

For this case, we use the pseudospin formalism to describe the spin dynamics of the polaritons in the cavity plane.
The pseudospin vector per particle $\textbf S(t)$ is a three-dimensional vector whose in-plane components describe the two linear polarization orientations of a given state, and whose normal-to-plane component $S_z$ is proportional to the circular polarization of the given polariton state. It fully corresponds to the Stokes polarization vector in the domain of classical optics.

The transverse-longitudinal splitting and the real magnetic field directed in the postive $z$ direction normal to the cavity plane give the reciprocal space effective Hamiltonian
\begin{equation}
H = \frac{\hbar^2 k^2}{2m^*} + (\boldsymbol\sigma\textrm{\textbf{H}}_{\textrm{eff}}),
\label{Hamilt}
\end{equation}
where the polariton effective mass is denoted by $m^*$, and we measure the effective magnetic field in energy units corresponding to splitting between orthogonal polarization components. The effective magnetic field represents a sum of the transverse-longitudinal field and the real magnetic field provoking the Zeeman splitting,

$$\textbf{H}_{\textrm{eff}}=\hbar\boldsymbol\Omega_{\textrm{\textbf{k}}}= \textbf{H}_{\textrm{LT}} + \textbf{H}_{\textrm{real}}$$

where the in- plane field $\textbf{H}_{\textrm{LT}}$ is given by Eq.\ref{HLT} and the z- directed field $\textbf{H}_{\textrm{real}}$ appears due to the application of the real magnetic field provoking a Zeeman splitting in circularly polarized components $\Delta_Z$. The components of the vector $\boldsymbol\Omega_{\textrm{\textbf{k}}}$ thus read
\begin{eqnarray}
\Omega_x = \frac{\Delta_{\textrm{LT}}}{2\hbar k^2}(k_x^2 - k_y^2),\label{OmegaX}\\
\Omega_y = \frac{\Delta_{\textrm{LT}}}{\hbar k^2}k_xk_y,\label{OmegaY}\\
\Omega_z=\frac{\Delta_Z} {2\hbar}\label{OmegaZ}.
\end{eqnarray}
with $\Delta_{\textrm{LT}}$ being the longitudinal-transverse splitting of the polariton doublet.
$\textbf{H}_{\textrm{real}} = H_{\textrm{real}} \hat{\boldsymbol{z}}$ is the real magnetic field.

The equation of motion for the scattered state pseudospin vector $\textbf{S}_{\textbf k}$ becomes;

\begin{equation} 
\frac {\partial \textbf{S}_{\textbf{k}}} {\partial t} = \textbf{S}_{\textbf{k}} \times \boldsymbol\Omega_\textbf{k} + \textbf{f}(t) - \frac{\textbf{S}_\textbf{k}} {\tau}
\label{PrecessionEq}
\end{equation}
with $\textbf{S}_{\textbf{k}}(0) = \textbf 0$.

The last term on the right-hand side of Eq.\ref{HLT} corresponds to the finite polariton lifetime $\tau$ and the second term stems from the flux of polaritons into the state ${\textbf{k}}$ due to the scattering from the state with $\textbf{k}=\textbf{k}_0$ where they were introduced at t=0 by an X- linearly polarized laser pulse. We consider the situation when the  scattering rate does not depend on scattering angle and model the flux by the expression: $$\textbf{f}(t) = \frac {\textbf{S}_0 (t)} {\tau_1} e^{-t/\tau},$$ where $\tau_1$ is the scattering time and $\tau$ is again the polariton lifetime. $\textbf{S}_0 (t)$ describes the evolution of the pseudospin in the pumped state. If the external magnetic field is absent and polaritons are pumped in the transverse or longitudinal linear polarization, then $\textbf{S}_0 (t)= const$. However, the presence of the magnetic field leads to the precession of $\textbf{S}_0 (t)$ around the z- axis before the scattering act, and its dynamics should be found from the following equation:
\begin{equation}
\frac {\partial \textbf{S}_0}{\partial t} = \textbf{S}_0 \times \boldsymbol\Omega_{0} 
\end{equation}
where $\textbf S_0 (0) = S_0\hat{\textbf x}$.

The circular polarization degree, $\rho _{c,\textbf{k}}$, can be defined by $$\rho_c \equiv 2S_{z,\textbf{k}}/N_\textbf{k}$$ where $N_\textbf{k}$ is the population of the scattered polariton state
which can be obtained from the following differential equation:
\begin{equation}
\frac {\partial N_\textbf{k}}{\partial t} = 2\frac{S_0}{\tau_1}e^{-t/\tau} - \frac{N_\textbf{k}}{\tau}
\end{equation}
where $S_0$ is the absolute value of $\textbf{S}_0(t)$, which gives
\begin{equation}
N_\textbf{k} = \frac{2 S_0 t}{\tau_1}e^{-t/\tau}.
\end{equation}

\begin{figure}
\centering

\includegraphics[width=3.41082in]{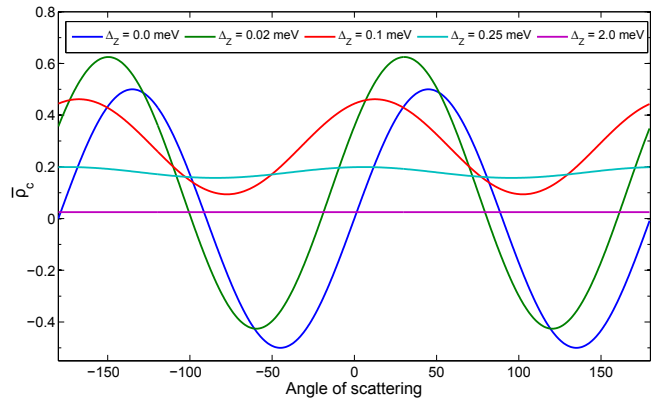}
\caption{Time-averaged circular polarization degree $\overline{\rho}_c$ plotted against the scattering angle for different magnitudes of the Zeeman-splitting $\Delta_Z$ indicated by different colors as shown in the legend on the top. For no external magnetic field, the polarization degree is devided into four quadrants which have alternating signs. As the field is increased, the pattern moves sideways and also upwards. For very strong external magnetic field, the azimuthal separation of polarizations fades out.}
\label{Time-average}
\end{figure}

Equation \ref{PrecessionEq} can be solved analytically (see Appendix). A plot of the time- averaged circular polarization as a function of the scattering angle is shown in Fig. \ref{Time-average} for several values of the external magnetic field. The lifetime of polaritons was taken to be $\tau = \hbar/\Delta_{LT}$ for which the maximum value of polarization occurs in the limit of vanishing external magnetic field. For vanishing $\Delta_Z$ the pattern is exactly the same as the one obtained by Kavokin et al \cite{Kavokin2005}. The increase of the field leads to the shift of the positions of the maxima and decrease of the amplitude of the oscillations, which gradually destroys the effect. The shift of the positions of the maxima corresponds to the rotation of the circular polarization degree pattern in the QW plane, which can also be seen in Fig. \ref{gaussian_circular}.

\subsection{Multiple scattering}

In the previous subsection we reached the conclusion that a strong magnetic field destroys  the OSHE. This was obtained with the use of a simplified model, based on the assumption that multiple scattering can be neglected and redistribution of the particles goes around the elastic circle only. In this section we show that relaxing these assumptions allows correction of this result and that magnetic field leads to the crossover from azimuthal separation of circular polarizations to radial separation of circular polarizations.

To investigate the effects of multiple scattering we resort to numerical modeling.  We consider the two-component macroscopic wavefunction of the polaritons in the circular polarization basis, 

\begin{eqnarray}
\boldsymbol\Psi =
\left(\begin{array}{c} \psi_+(\textbf{r},t) \\ \psi_-(\textbf{r}, t) \end{array}\right)
\end{eqnarray}
where $\psi_+$ represents the right circularly polarized component and $\psi_-$ the left circularly polarized component. Knowing $\boldsymbol\Psi$ allows straightforward determination of the Stokes parameters of the emitted light, corresponding to circular polarization  degree $\rho_c$ and polarization degrees in the linear ($\rho_{lin}$) and diagonal components ($\rho_{diag}$)\cite{ShelykhKavokin2010}
\begin{eqnarray}
\rho_c = \frac{|\psi_+|^2 - |\psi_-|^2}{|\psi_+|^2 + |\psi_-|^2},\label{rhoc}\\
\rho_{lin}  = \frac{\psi_+^*\psi_- + \psi_-^*\psi_+}{|\psi_+|^2 + |\psi_-|^2},\label{rholin}\\
\rho_{diag} = i\frac{\psi_+^*\psi_- - \psi_-^*\psi_+}{|\psi_+|^2 + |\psi_-|^2}.\label{rhodiag}
\end{eqnarray}
The macroscopic wavefunction satisfies the Schr\"{o}dinger- type equation accounting for the external pump and decay:
\begin{equation}
i\hbar\frac {\partial \boldsymbol\Psi}{\partial t} =-i\hbar\frac{\boldsymbol\Psi}{2\tau}+ \boldsymbol \hat H \boldsymbol \Psi + \boldsymbol \hat V(\textbf r)\boldsymbol \Psi + \boldsymbol P(\textbf{r},t)
\label{Schrodinger}
\end{equation}

The first term corresponds to the finite life of the polaritons.

The second term describes the in- plane propagation of free polaritons and rotation of their pseudospin under the effect of the effective magnetic field. The corresponding Hamiltonian reads
\begin{eqnarray}
\boldsymbol \hat H = \left(\begin{array}{cc} \frac{-\hbar^2\nabla^2}{2m}+\frac{\Delta_Z}{2} & \beta\left(\frac{\partial}{\partial x} + i\frac{\partial}{\partial y}\right)^2 \\ \beta\left(\frac{\partial}{\partial x} - i\frac{\partial}{\partial y}\right)^2 &  \frac{-\hbar^2\nabla^2}{2m}-\frac{\Delta_Z}{2}  \end{array}\right)
\end{eqnarray}
where the diagonal elements corresponds to the kinetic energy and Zeeman splitting of the polaritons and the off-diagonal elements represent the TE-TM splitting. This expression can be straightforwardly obtained from Eqs.\ref{Hamilt},\ref{OmegaX},\ref{OmegaY},\ref{OmegaZ} using the substitution $\hat{k}\rightarrow-i\hbar\nabla$.

The term $\hat V$ represents the external scattering potential for which we use a model of Gaussian correlated disorder \cite{Savona2006}. Both the excitonic and photonic parts were considered. The excitonic and photonic correlation lengths of the disorder potential were taken to be $\sigma_X = 0.3$ $\mu$m and $\sigma_C = 1$ $\mu$m, respectively. The depth of the potentials were $A_X = 0.15$meV for the excitonic part and $A_C = 0.4$meV for the photonic part. These parameters were taken from the work of T. C. H. Liew \textit{et al.}\cite{Liew2009}.

The last term correspond to the external coherent pump of the system, which in our case represents a stationary homogeneous x-directed linearly polarized optical field which has an in-plane wave-number $k_0 = 1.56$ $ \mu m^{-1}$ corresponding to the value of TE-TM splitting of $\Delta_{LT} = 0.05$ meV.

\begin{figure*}
\includegraphics[width=\textwidth]{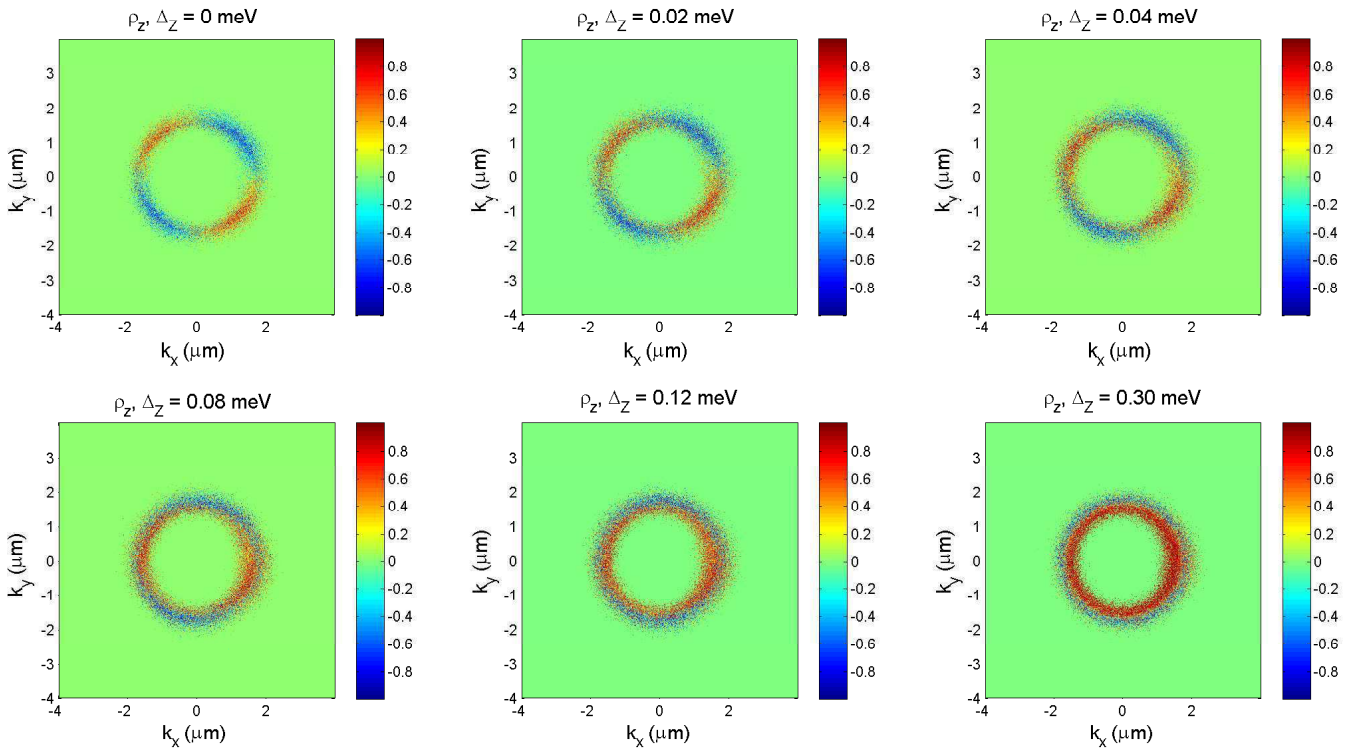}
\caption{Numerical calculation of the evolution of the steady-state circular polarization degree in reciprocal space with increasing strength of Zeeman-splitting $\Delta_Z$ as shown above each subplot. The homogeneous CW pump has an in-plane wave-vector $\boldsymbol k_0$ in the positive x direction and multiple scattering with a disorder potential is accounted for. It is visible how the four-leaf pattern transforms into a separation of polarization in the radial direction with increasing external magnetic field.}
\label{multiscattering}

\end{figure*}

The results of numerical modeling corresponding to the stationary regime are shown in Fig. \ref{multiscattering}. The figure illustrates the distribution of the circular polarization degree in reciprocal space. With no external field applied one clearly sees formation of the domains with alternating positive and negative circular polarization degree in the four half- quarters of the XY plane. The increase of the magnetic field leads to the slight rotation of these domains until $\rho_c$ becomes azimuthally symmetric, which corresponds to the result obtained in the framework of the simplified model with a single scattering event. Further increase of the magnetic field leads to the radial separation of the circular polarizations. This effect can be qualitatively explained as follows. In the regime when Zeeman splitting dominates over TE- TM splitting, the eigenstates of the Hamiltonian become circularly polarized. The states having opposite circular polarization corresponding to the same pump energy $\hbar\omega_P$ correspond to different wavenumbers, $k_\pm=\sqrt{2m(\hbar\omega_P\mp\Delta_Z/2)/\hbar^2}$ (see Fig. \ref{scheme}). Therefore the energy- conserving elastic scattering goes preferably along different elastic circles separated by the distance
\begin{equation}
\Delta k \approx\Delta_Z\sqrt{\frac{m}{2\hbar^3\omega_P}}
\end{equation}
For not very strong fields this value is expected to be small in GaAs based cavities, but can be sufficiently increased in cavities based on materials with large g- factors (e.g. InGaAs) or in semimagnetic cavities\cite{Shelykh2009b}. 

\section{Geometry of the spatially localized pump}
As already mentioned, the alternative geometry for the observation of the OSHE consists in pumping of the disorder- free cavity with spatially localized spot. The effects of longitudinal-transverse splitting manifest themselves in the formation of cross- like polarization patterns in the near- field emission, reported in Refs.\onlinecite{Kammann2012,Amo2009}. In this section we investigate how these patterns change under the effect of the external z- directed magnetic field.

For modeling of the dynamics of the polaritons we again use the equation \ref{Schrodinger}, putting $\hat{V}(\textbf{r})=0$ and modeling the pump, focused at the the center of the quantum well as  $$ \boldsymbol P(\textbf r,t) = A_0 e^{-\frac{iE_pt}{\hbar}} e^{-\frac{r^2}{r_0^2}}\left(\begin{array}{c} 1 \\ 1  \end{array}\right)$$
where $E_p = \hbar\omega_P$ is the energy of the pump relative to the minimum of the lower polariton branch. This pump is switched on at the time $t=0$. 
According to Eqs.\ref{rhoc},\ref{rholin},\ref{rhodiag} the determination of the macroscopic wavefunction $\boldsymbol\Psi = (\psi_+,\psi_-)^T$ allows finding the Stokes parameters of the system corresponding to circular and linear polarization degrees and connected to components of the pseudospin $S_x,S_y$ and $S_z$. 

\begin{figure*}
\includegraphics[width=\textwidth]{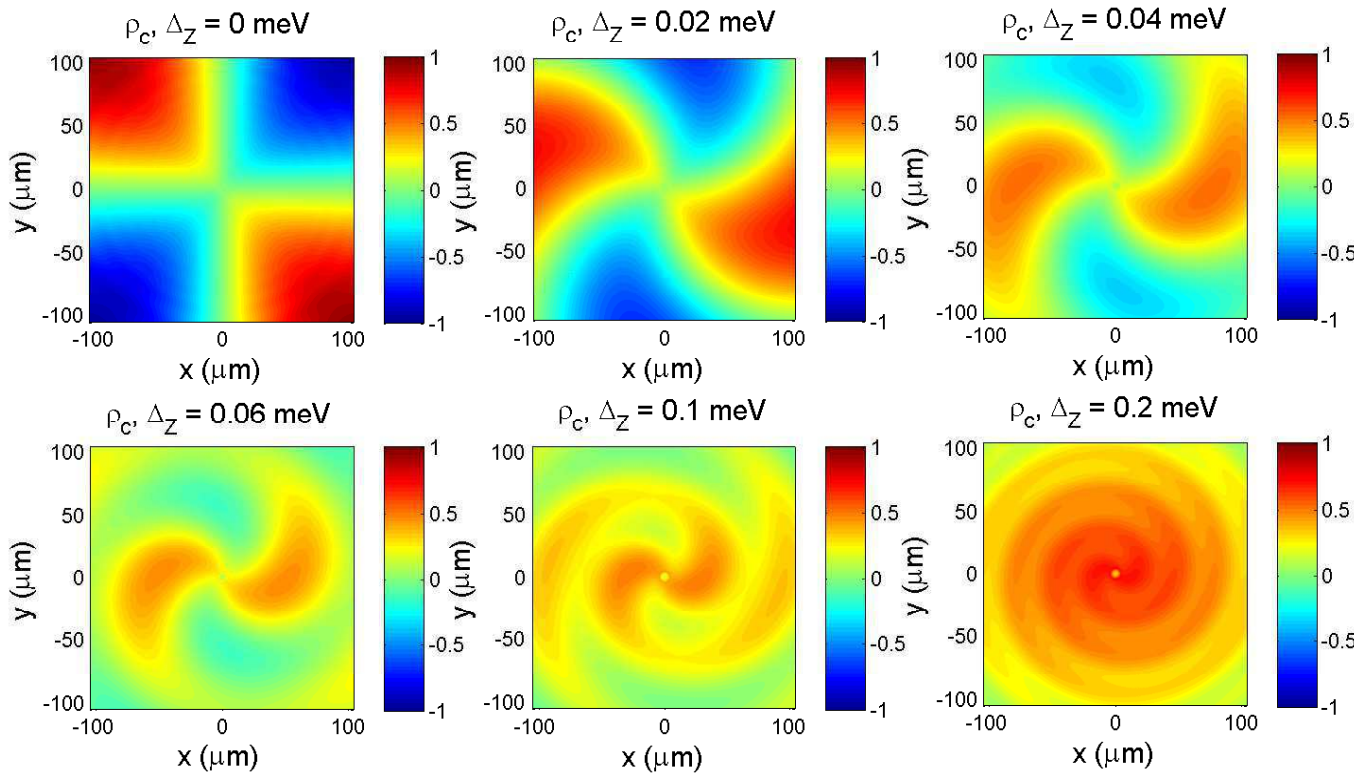}
\caption{The dependence of the steady-state real space circular polarization degree $\rho_x$ on the real magnetic field directed out of the plane. In this case, the pump is a Gaussian CW pump normal to the plane with a spot size of 3 $\mu$m and linear polarization. For no magnetic field, we see a cross-like separation of circular polarizations but as the field is slightly increased, the cross begins to rotate and forms a gammadional shape. For higher magnetic fields, the polarization pattern rotates still further and starts to form a a spiral and also decreases in magnitude.}
\label{gaussian_circular}
\end{figure*}

The steady-state pattern of the circular polarization degree, $\rho_c$, is shown in 
Fig. \ref{gaussian_circular}. When there is no external magnetic field applied, we see the usual cross pattern characteristic of the OSHE. As the field is increased by a small value, the cross starts to rotate and the circular polarization degree makes a gammadion-like shape. For higher magnetic fields the arms of the cross start to form spirals centered at the pumping spot. The maxima of the absolute value of the polarization degree start to decrease in this case.

\begin{figure*}

\includegraphics[width=\textwidth]{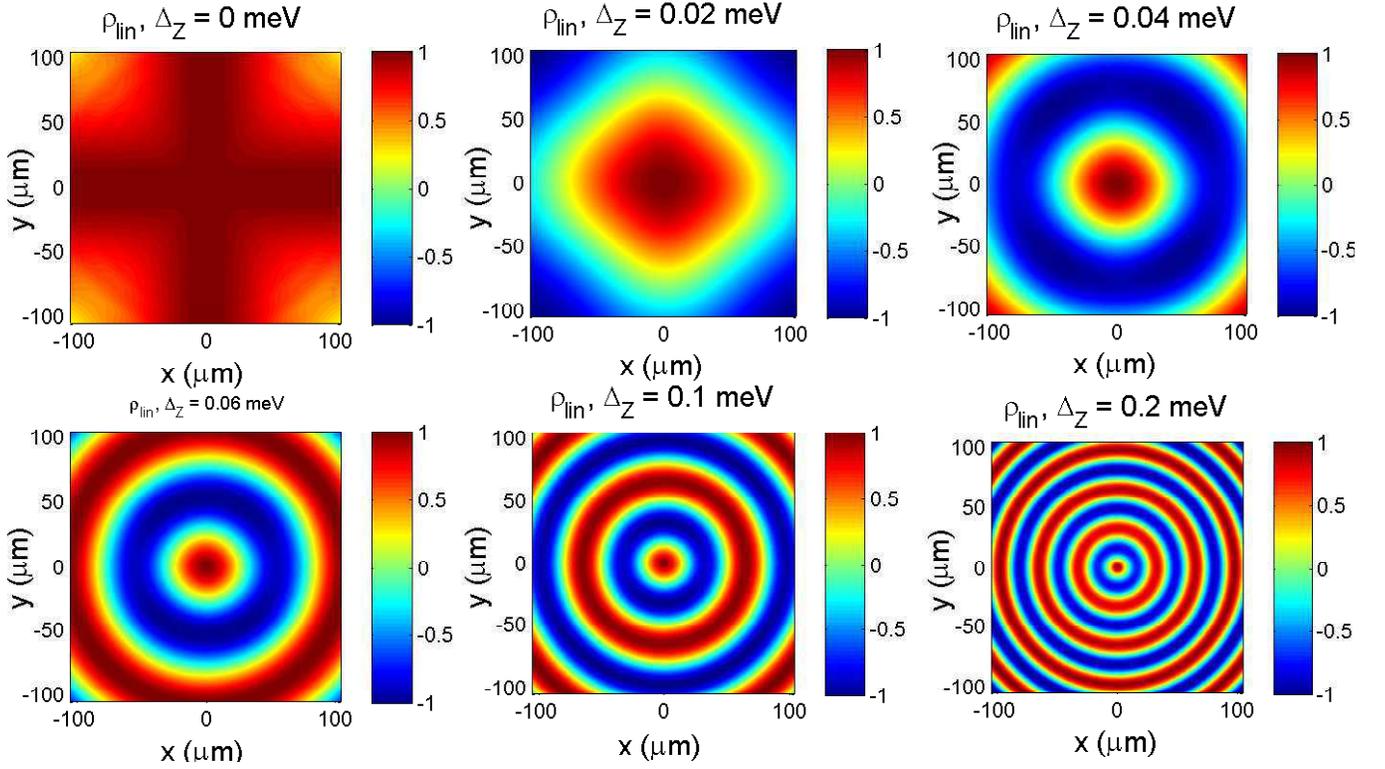}
\caption{The dependence of the steady-state real space linear polarization degree $\rho_{lin}$ on the real magnetic field directed out of the plane. In this case, the pump is a Gaussian CW pump normal to the plane with a spot size of 3 $\mu$m and linear polarization.  For no external magnetic field , the linear polarization degree makes a cross-like pattern. For a slight increase in applied field, the patter becomes a rounded rectangle and as the field is further increased, we observe the formation of rings centered at the pumping spot. The difference in the radii of the rings decreases with increasing external magnetic field which corresponds to an increase in rotation frequency of the polariton pseudospin vector around the total effective magnetic field.}
\label{gaussian_linear}

\end{figure*}

The pattern of the linear polarization degree, $\rho_{lin}$, is shown in Fig. \ref{gaussian_linear}. For vanishing external field, we observe a distribution of linearly polarized polaritons along the x- and y-axes. When the applied field is increased by a small value, a rounded rectangle is formed inside which the linear polarization of polaritons is of different sign those outside of the rectangle. As the field is further increased, we observe formation of rings centered at the pumping spot. The separation of the rings decreases with increasing magnetic field. 

\begin{figure*}

\includegraphics[width=\textwidth]{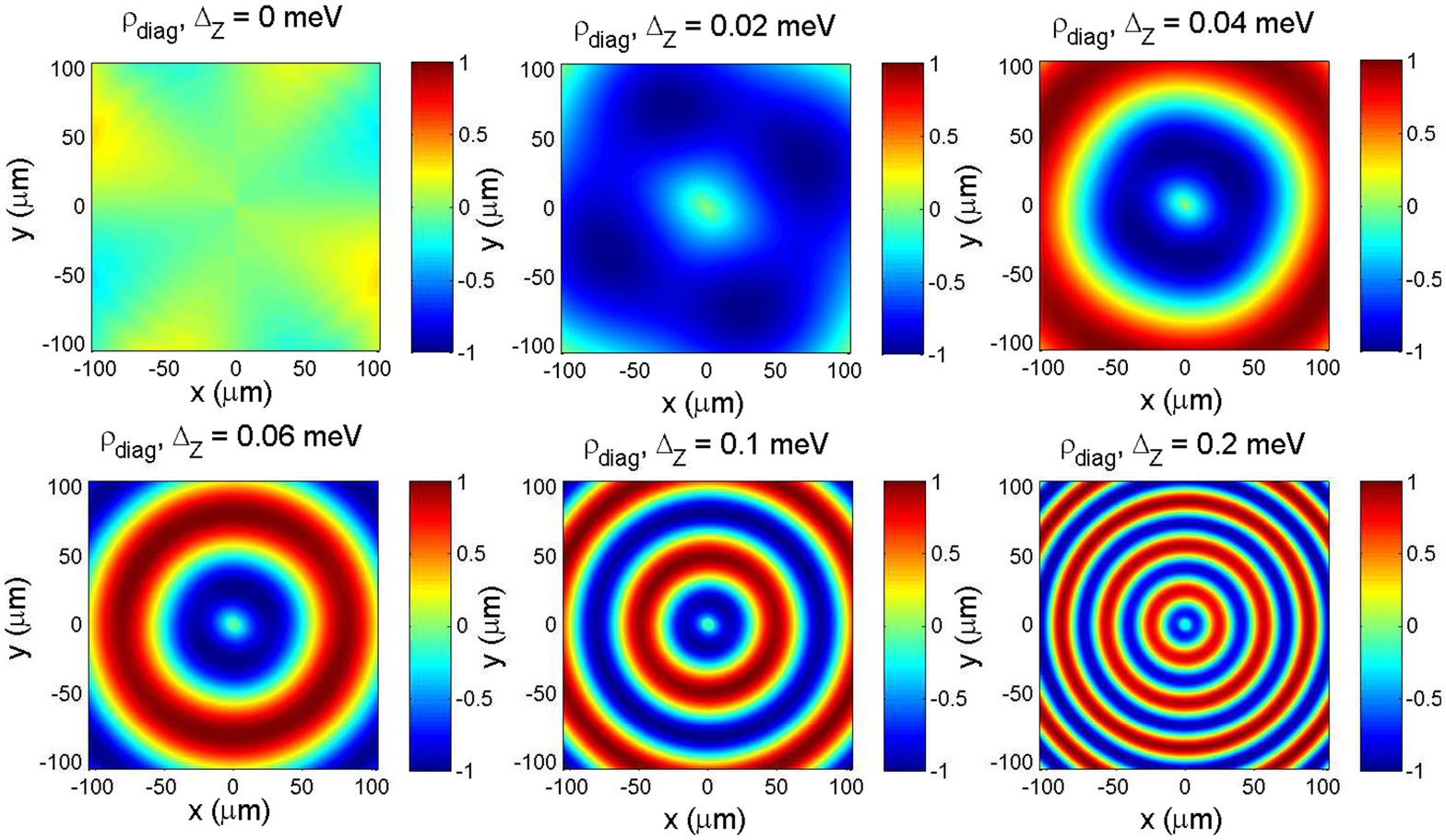}
\caption{The dependence of the steady-state real space diagonal polarization degree $\rho_{diag}$ on the real magnetic field directed out of the plane. In this case, the pump is a Gaussian CW pump normal to the plane with a spot size of 3 $\mu$m and linear polarization.  As for the linear polarization degree, $\rho_{diag}$ forms a cross-like pattern for no external magnetic field. As the field is increased, a rounded rectangle forms. For higher magnetic fields, rings centered at the pumping spot occur and their separation decreases with increasing magnetic field which corresponds to an increase in the rotation frequency of the pseudospin vector of polaritons around the total effective magnetic field. }
\label{gaussian_diagonal}

\end{figure*}

In the same way, the diagonal polarization degree of the polaritons, $\rho_{diag}$, forms a pattern which is split into four quadrants as shown in Fig. \ref{gaussian_diagonal}. As the field is slightly increased, $\rho_{diag}$ forms a rounded square around the pumping spot, but when the applied field is higher, rings centered at the pumping spot occur. The difference in radii of adjacent rings decreases with increasing external magnetic field.

We can thus explain the results of Figs. \ref{gaussian_linear}  and \ref{gaussian_diagonal} by noting that when polaritons move radially from the the center, the pseudospin rotates around the effective magnetic field which lies in the XY-plane for vanishing applied magnetic field. This gives us the separation of circular porization into quadrants. When the applied magnetic field is much higher than the field arising from the LT-splitting of polaritons, the effective total magnetic field is parallel to the z axis and since we are pumping with a linearly polarized light, the pseudospin rotates around the z axis in the XY plane. The increase in the applied field gives a higher rotation frequency and this manifests itself in the formation of rings in the linear and diagonal polarization degrees' patterns.


\section{Conclusion}
We analyzed the effect of an applied magnetic field on the optical spin Hall effect in semiconductor microcavities in the geometries of the disordered cavity excited by homogenous laser beam and clean cavity excited by localized laser spot. We showed that in both cases magnetic field dramatically changes the pattern of spin currents. For the disordered cavity the most basic single scattering semiclassical approximation predicts suppression of spin currents in a strong magnetic field. Account for multiple scattering leads to the transition for azimuthal to radial separation of circular polarizations in the reciprocal space. For tightly focused laser excitation the magnetic field leads to the warping of the pattern of circular polarization in the real space, changing it from cross- like to gammadion- like. It also causes the formation of a series of concentric linearly polarized rings in real space.

We thank Prof. Yu. Rubo for the discussions we had on the topic of the paper. The work was supported by FP7 IRSES projects "SPINMET" and "POLAPHEN" and Tier 1 project "Polaritonics for Novel Device Applications". S.M. thanks Universidad Autonoma de Mexico for the hospitality.

\newpage
\section{Appendix}

To solve the initial value problem around Eq.\ref{PrecessionEq} we make the substitution $\textbf{S}'_{\textbf{k}}(t) = \textbf{S}_{\textbf{k}}e^{t/\tau}$
and obtain the simplified problem

$$\frac {\partial \textbf{S}'_{\textbf{k}}(t)}{\partial t} = \textbf{S}'_{\textbf{k}}(t) \times \boldsymbol{\Omega} + \frac{\textbf{S}_{\textbf{0}}(t)}{\tau_1} = -i\hat A \textbf{S}'_{\textbf{k}}(t) + \frac{\textbf{S}_{\textbf{0}}(t)}{\tau_1}$$
where $\hat{A}$ is a Hermitian 3x3 matrix and $\textbf{S}'_{\textbf{k}}(0) = \textbf 0$.

The matrix $\hat A$ has three real eigenvalues, $\omega_1, \omega_2, \omega_3$, with three corresponding eigenvectors $\textbf{e}_1,\textbf{e}_2,\textbf{e}_3$, respectively, which form a basis on the 3-D Euclidian space. Thus, we can write
$$\textbf{S}'_{\textbf{k}}(t) = c_1(t)\textbf{e}_1 + c_2(t)\textbf{e}_2 + c_3(t)\textbf{e}_3$$
and
$$\frac{\textbf{S}_{\textbf{0}}(t)}{\tau_1} = g_1(t)\textbf{e}_1 + g_2(t)\textbf{e}_2 + g_3(t)\textbf{e}_3.$$

From this, we get three uncoupled ordinary differential equations,

\begin{eqnarray}
c_j'(t) &=& -i\omega_j c_j(t) + g_j(t) \\
\Rightarrow
 c_j(t) &=& e^{-i\omega_j t}\int_0^t{ e^{i\omega_j t'}g_j(t')dt'},
\end{eqnarray}

with $c_j = 0$ at $t=0$ for $j=1,2,3$.
These equations are solved straightforwardly to give the time-averaged circular polarization value

\begin{widetext}
\begin{align}
\notag
\overline\rho_c(\phi) =& \: 2\frac{\int_0^\infty S_{\boldsymbol k,z}dt}{\int_0^\infty N_{\boldsymbol k}dt}\\
\notag
 =& \: \frac 1 {\Omega^2}\left[\Omega_x \cos^2\theta + \Omega_z\cos \theta\sin\theta\right]\Omega_z\\
\notag
 &+ \Omega_z\left[\Omega_x-\Omega\ \cos\theta\right](3\sin^2\theta-1)\frac{1}{2((\Omega\tau)^2+1)}\frac 1 {\Omega^2}\\
\notag
 &+ \cos^2\theta\frac{\Omega_y\tau}{(\Omega\tau)^2 + 1} + 2 \sin\theta   \frac{\Omega_y\Omega_z\tau}{(\Omega\tau)^2 + 1)^2}\frac 1 {\Omega}\\
 &-\left[\Omega_x\Omega_z(1+\sin^2\theta) + \Omega^2\cos^3\theta\sin\theta\right]\frac{(1 - \Omega^2\tau^2)}{((\Omega\tau)^2+1)^2}\frac 1 {2\Omega^2}.
\end{align}

where $\theta$ is the angle the effective magnetic field makes with the cavity plane, given by $\tan(\theta) = \frac{\mu_B g H_{\textrm{real}}}{\Delta_\textrm{LT}}$.

\end{widetext}

\end{document}